\documentclass[prb,preprint,aps,superscriptaddress,longbibliography]{revtex4-1}

\usepackage{graphicx, epstopdf}
\usepackage{subfigure}
\usepackage{float}
\usepackage{amsmath}
\usepackage{amsfonts}
\usepackage{amssymb}
\usepackage{bm}
\usepackage{subfigure}
\usepackage{caption}
\usepackage{hyperref}
\begin{document}

\title{Co-existence of Weyl Fermion and Massless Triply Degenerate Nodal Points}


\author{Hongming Weng}
\email{hmweng@iphy.ac.cn}
\affiliation{Beijing National Laboratory for Condensed Matter Physics,
  and Institute of Physics, Chinese Academy of Sciences, Beijing
  100190, China}
\affiliation{Collaborative Innovation Center of Quantum Matter,
  Beijing, China}

\author{Chen Fang}
\email{cfang@iphy.ac.cn}
\affiliation{Beijing National Laboratory for Condensed Matter Physics,
  and Institute of Physics, Chinese Academy of Sciences, Beijing
  100190, China}

\author{Zhong Fang}
\affiliation{Beijing National Laboratory for Condensed Matter Physics,
  and Institute of Physics, Chinese Academy of Sciences, Beijing
  100190, China}

\affiliation{Collaborative Innovation Center of Quantum Matter, Beijing, China}

\author{Xi Dai}

\affiliation{Beijing National Laboratory for Condensed Matter Physics,
  and Institute of Physics, Chinese Academy of Sciences, Beijing
  100190, China}

\affiliation{Collaborative Innovation Center of Quantum Matter, Beijing, China}

\date{\today}

\begin{abstract}
By using first-principles calculations, we propose that WC-type ZrTe is a new type of topological semimetal (TSM).
It has six pairs of chiral Weyl nodes in its first Brillouin zone, but it is distinguished from other existing TSMs by having additional two paris of massless fermions with triply degenerate nodal points as proposed in the isostructural compounds TaN and NbN. 
The mirror symmetry, three-fold rotational symmetry and time-reversal symmetry require all of the Weyl nodes to have the same velocity vectors and locate at the same energy level. 
The Fermi arcs on different surfaces are shown, which may be measured by future experiments. 
It demonstrates that the ``material universe" can support more intriguing particles simultaneously.
\end{abstract}

\maketitle

\section{Introduction} \label{introduction}

Different types of topological semimetals (TSMs) have been proposed and studied intensively in recent 
years.~\cite{TSM_review,Chiu_RMP,Bansil_RMP,WSM_review} All these TSMs can be classified according 
to the different band crossing points at the Fermi level and the mechanism protecting them. For example, Dirac 
semimetal (DSM) ~\cite{young_dirac_2012, Na3Bi, Cd3As2} can be either protected by special crystalline 
symmetries on high symmetrical crystal momentum points or along high symmetry lines,~\cite{yang_classification_2014} 
and Weyl semimetal (WSM) ~\cite{murakami_phase_2007,wan,HgCrSe,TaAs_Weng,HuangSM_Weyl,Lu622,TaAs_arc, TaAs_node, Xu2015a, Yang2015,Weyl_arc_ossi,Soluyanov2015} can be stabilized without any additional crystalline symmetry 
besides the lattice translation. Exotic surface states forming open ``Fermi arcs"~\cite{wan,HgCrSe,Na3Bi} 
and negative magnetoresistance~\cite{Son2013,TaAs_anomaly, Transport_Weyl_XLQi_2013} caused by ``chiral anomaly" 
are two important physical consequences of Dirac and Weyl semimetal states. Very recently, another class 
of TSM materials characterized by three or six fold band crossings was proposed as having ``New Fermions"
with quite interesting physical properties been discussed in References~\onlinecite{newfermion,InAsSb_TDNP,TaN_TDP}. 

In the present paper, we propose that ZrTe is a material that in one hosts both the ``New Fermion" state with triply degenerate 
crossing points and the Weyl Fermion state near the Fermi energy. ZrTe has the same crystal structure and similar band structure along 
the $\Gamma$-A direction with TaN, which has triply degenerate nodes.~\cite{TaN_TDP} While what is new in ZrTe is that at the same time the 
band structure of ZrTe also contains six pairs of Weyl points near the mirror plane of $k_z$=0. The microscopic mechanism 
for the formation of the Weyl points in ZrTe is similar to TaAs,~\cite{TaAs_Weng} where the non spin-orbit coupled (SOC) band structure contains 
nodal lines on the mirror plane. Turning on SOC leads to the disappearance of the nodal lines on the mirror plane and the appearance of Weyl 
nodes off the plane. Compare to TaAs, an important advantage for the WSM phase in ZrTe is that all the Weyl points are 
related by crystalline symmetry $D_{3h}$, so that all of them are located at the same energy. Regarding the surface states, an interesting point in 
ZrTe as a WSM is that there are four different symmetric Fermi arc patterns allowed by symmetry on the (001) surfaces 
and fruitful surface Lifshitz transitions may be induced by changing the surface condition accordingly.

\section{Computational Details} \label{method}
The software package OpenMX~\cite{openmx} is employed for most of the first-principles calculations, which is based on the linear combination
of psuedo-atomic orbitals.~\cite{lcao_ozaki} The local orbital basis sets for Zr and Te are chosen as Zr9.0-s2p2d2f1 and Te7.0-s2p2d1, respectively. 
The experimental crystal structure of WC-type ZrTe has been studied by \"{O}rlygsson {\it et al.}~\cite{WC-ZrTe, MnP-ZrTe} and is fully relaxed 
until the residual forces on each atom is less than 0.001 eV/\AA. The sampling of the Brillouin zone in the self-consistent process is taken as the grid of
12$\times$12$\times$10. Exchange-correlation potential is treated within the generalized gradient approximation (GGA) of Perdew-Burke-Ernzerhof 
type.\cite{Perdew1996} The possible underestimation of band gap within GGA is checked by non-local Heyd-Scuseria-Ernzerhof (HSE06) hybrid 
functional\cite{heyd2003hybrid, heyd2006hybrid} calculation using VASP software package.~\cite{kresse1996_1, kresse1996_2} Spin-orbit coupling (SOC) 
is taken into account self-consistently in both software packages. The maximally localized Wannier functions (MLWF)~\cite{marzari1997,souza2001} 
for the $d$ orbitals of Zr are generated by using OpenMX.~\cite{weng_mlwf} Then, a tight-binding model based on these has been established to 
calculate the surface states with the Green's function methodology.~\cite{MRS_weng:9383312, adv_phys} 

\begin{figure}
\includegraphics[width=0.8\textwidth]{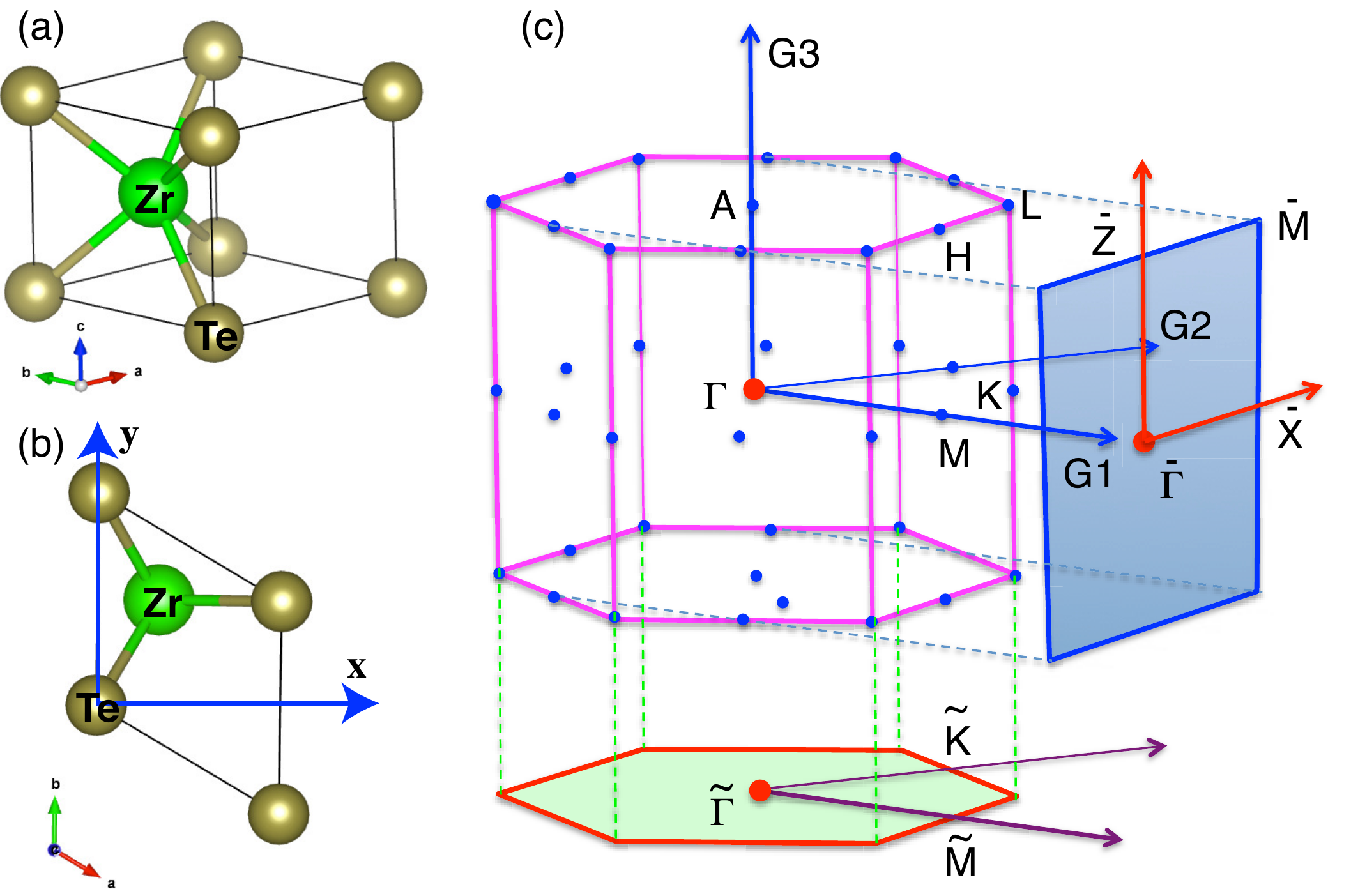}
\caption{(Color online)  (a) The crystal structure of WC-ZrTe and its (b) the top view. (c) The bulk Brillouin zone and its projection onto the (100) and (001) surface.
The high symmetrical crystal momenta are indicated. 
}
\label{crystructure}
\end{figure}

\section{Results and Discussion} \label{Results}
{\it Crystal Structure.} 
As shown in Fig.~\ref{crystructure}, the WC-type ZrTe has the same crystal structure as $\theta$-TaN~\cite{thetaTaN,thetaTaN-2} and NbN.~\cite{NbN_1989,NbN_1990} 
Hereafter, we will simply call them ZrTe, TaN and NbN, respectively. The space group is $P\bar{6}m2$ (No. 187). Zr and Te atoms occupy the 1$d$ (1/3, 2/3, 1/2) and 1$a$ (0,0,0) 
Wyckoff positions, respectively. The experimental lattice constants are $a$=$b$=3.7707~\AA~ and $c$=3.8606~\AA. The optimized lattice 
constants are $a$=$b$=3.7966~\AA~ and $c$=3.9003~\AA, being slightly overestimated by 1.0\%. All the results discussed in the following are
from the calculations with optimized ones.
 
\begin{figure}
\includegraphics[width=0.8\textwidth]{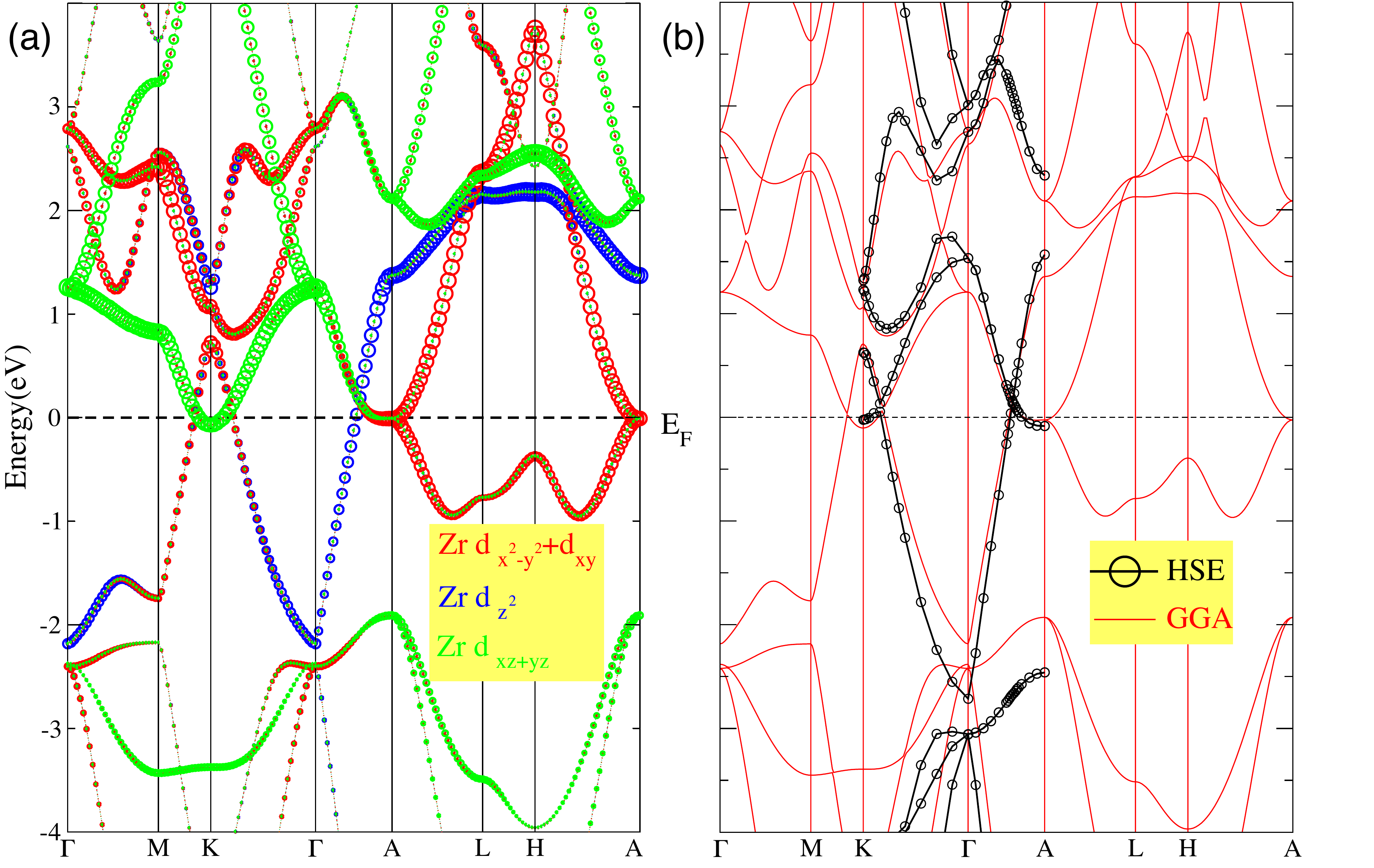}
\caption{(Color online) Band structure of ZrTe without SOC. (a) The fatted bands with spectral weight of Zr $d_{z^2}$, $d_{x^2-y^2}$+$d_{xy}$ and 
$d_{xz}+d_{yz}$ orbitals. (b) The band structures calculated within GGA (lines) and hybrid functional HSE06 (circled lines). 
}
\label{ZrTe_GGAband}
\end{figure}

{\it Band structure of ZrTe.} 
The band structure calculated within GGA is shown in Fig.~\ref{ZrTe_GGAband}. When compared with that of TaN~\cite{TaN_TDP},
the band crossings along $\Gamma$-A between $d_{z^2}$ and $d_{x^2-y^2}$+$d_{xy}$ bands are the same. The energy dispersions
along A-L-H-A within $k_z$=$\pi$ plane are also the same. The most distinct difference is within the $k_z$=0 plane , where 
additional band inversion around K happens in ZrTe but is absent in TaN. Such band inversion includes $d_{x^2-y^2}$+$d_{xy}$ and
$d_{xz}$+$d_{yz}$ bands. Since $k_z$=0 plane is a mirror plane and the two bands have opposite mirror eigenvalue, the band 
inversion leads to a circle of node-line centering K point, which is similar to the case in TaAs.~\cite{TaAs_Weng}
The hybrid functional calculation is used to overcome the possible overestimation of band inversion in GGA. The comparison of them
is shown in Fig.~\ref{ZrTe_GGAband}(b), which indicates that these two band inversions occur in both cases.

\begin{figure}
\includegraphics[width=0.8\textwidth]{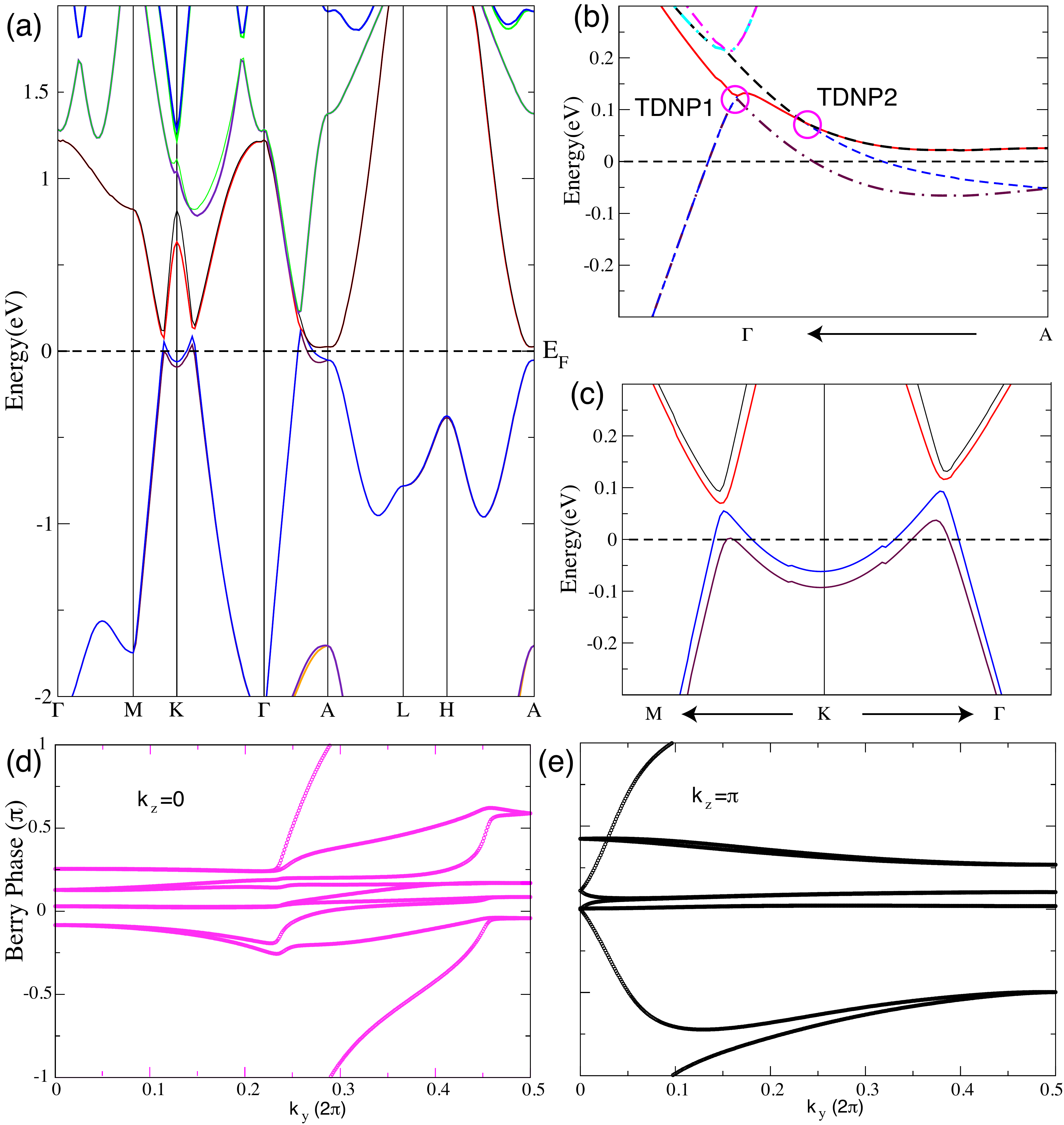}
\caption{(Color online) (a) Band structure with SOC included. (b) Enlarged band structure along M-K-$\Gamma$ and (c) $\Gamma$-A in (a). (d) The $k_y$ evaluation of Berry phases of occupied bands in $k_x$ periodicity in $k_z$=0 and (ed) $k_z$=$\pi$ (lower) plane for ZrTe. Both planes have nontrivial $Z_2$ number of 1. 
}
\label{ZrTe_socBand}
\end{figure}

With SOC being included, the bands in Fig.~\ref{ZrTe_socBand}(a) and (b) along $\Gamma$-A cross at two triply degenerate nodal points (TDNPs), which 
are the massless fermions discussed in TaN and NbN.~\cite{TaN_TDP} However, the additional band inversion around K results in six 
pairs of Weyl nodes in its first Brillouin zone (BZ). Since the properties associated with the TDNP have already been discussed in detail in
our previous paper on TaN and NbN, in this work we mainly focused on the properties related to the existence of Weyl points, 
which is absent in TaN or NbN. 

\begin{figure}
\includegraphics[width=0.8\textwidth]{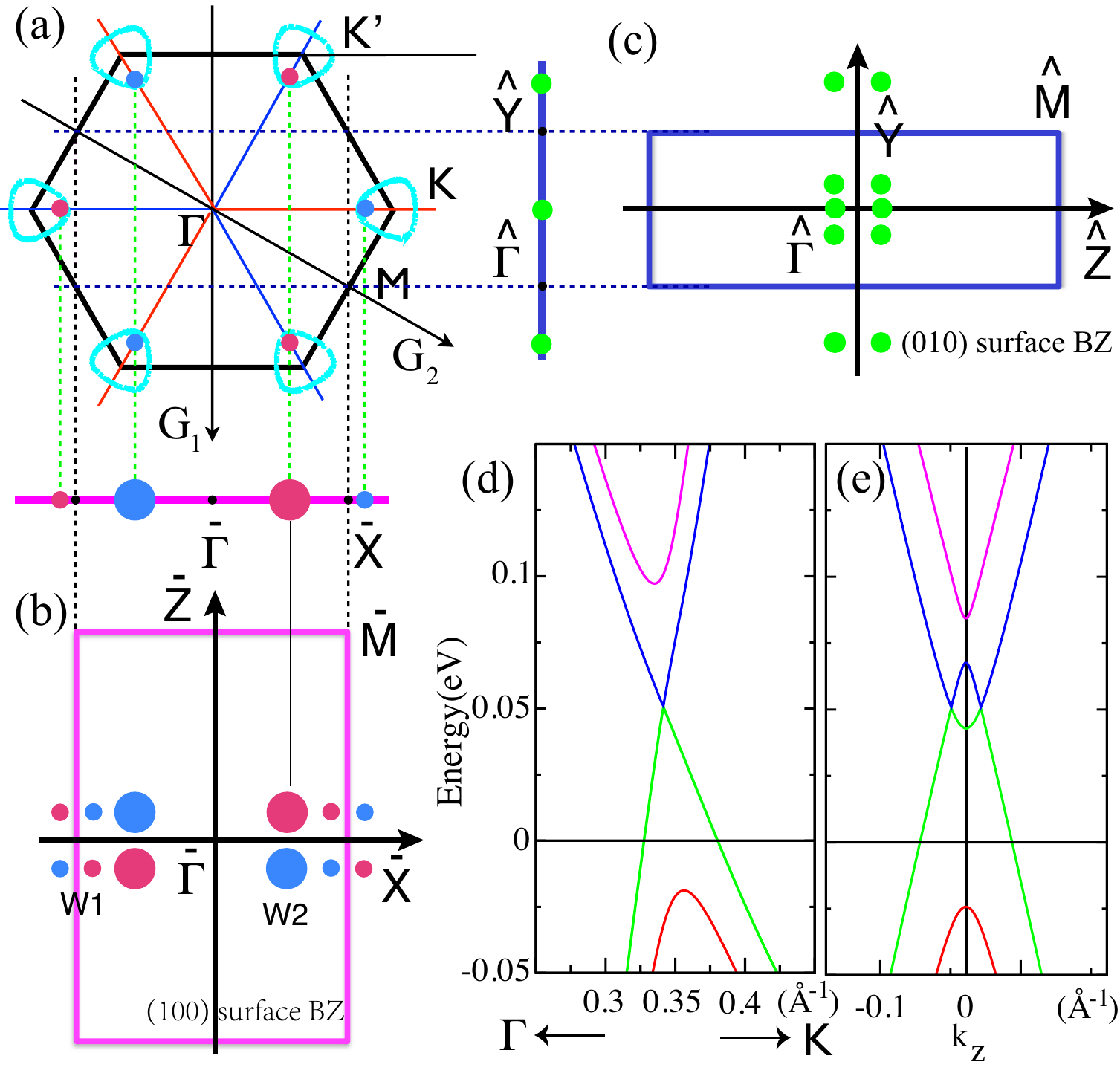}
\caption{
(Color online) (a) The node-lines centering K and K' in non-SOC case and they decay into Weyl nodes when SOC is included. (b) and (c) are their 
projections onto (100) and (010) surface, respectively. (d) the band structures around Weyl nodes. All of the Weyl nodes are related by symmetrical 
operation and have the same velocity matrix and the same nodal energy about 50 meV.
}
\label{ZrTe_Weyl_Dist}
\end{figure}

As shown in Fig.~\ref{ZrTe_socBand} (c), the lack of inversion symmetry leads to Rashba-like spin splitting in the bands along M-K-$\Gamma$.
The node-lines around both K and K' (time reversal partner of K) within GGA in Fig.~\ref{ZrTe_Weyl_Dist} (a) are fully gaped, but 
each of them will lead to three pairs of Weyl nodes off the $k_z$=0 mirror plane and will be discussed later. Since the electronic states in both the 
$k_z$=0 and $k_z$=$\pi$ planes can be looked as two-dimensional (2D) insulators with time-reversal and mirror symmetry, the band topology 
can be identified by their mirror Chern numbers (MCN) and $Z_2$ numbers. The wilson
loop method~\cite{YuRui_Z2_2011PRB, MRS_weng:9383312, adv_phys} is used to calculate the evolution of the ``Wannier centers''  formed 
along the $k_y$ direction, showing that at both planes, the MCN is one and $Z_2$ number odd [see Fig.~\ref{ZrTe_socBand}(d, e)]. 
Compared to the situation in TaN, we note that here both the $k_z$=0 and $k_z$=$\pi$ planes are topologically non-trivial, whereas in TaN 
only the $k_z=\pi$ plane is non-trivial. 

{\it Weyl nodes in ZrTe.} 
By first principle calculations, we have located the positions of all the WPs in $ZrTe$, which are shown in Fig.~\ref{ZrTe_Weyl_Dist}(a)
from the top view along the c-axis. Totally there are six pairs of WPs in the BZ with half of them being above and the other half being below
the $k_z=0$ plane. Due to the high crystalline symmetry in $ZrTe$, all the WPs can be related to each other by the $C_3$ rotation, the vertical 
mirror $\Gamma MA$ and the horizontal mirror $\Gamma MK$. Therefore all the WPs are located at the same energy 
similar to the situation in strained HgTe~\cite{HgTe-haijun} or chalcopyrites.~\cite{huangtong-haijun}.
One typical pair of WPs are located at (0.26984, 0.26984, $\pm$ 0.00655) in unit of reciprocal lattice vectors. 

When these Weyl nodes are projected onto (100) surfaces, the different chiral Weyl nodes are projected onto different position as shown
in Fig.~\ref{ZrTe_Weyl_Dist}(b) and there are two pairs superposed each other indicated with bigger circles labeled as W2. Others are W1.
When they are projected onto (010) surface, the Weyl nodes with opposite chirality are superposed as shown in Fig.~\ref{ZrTe_Weyl_Dist}(c), 
which is similar for the (001) surface. The Dirac cone like band structures passing through one of the Weyl node is plotted along 
$\Gamma$-K and $z$-axis in Fig.~\ref{ZrTe_Weyl_Dist}(d) and (e), respectively. The Weyl node is located at 50 meV above the Fermi level.

\begin{figure}
\includegraphics[width=0.8\textwidth]{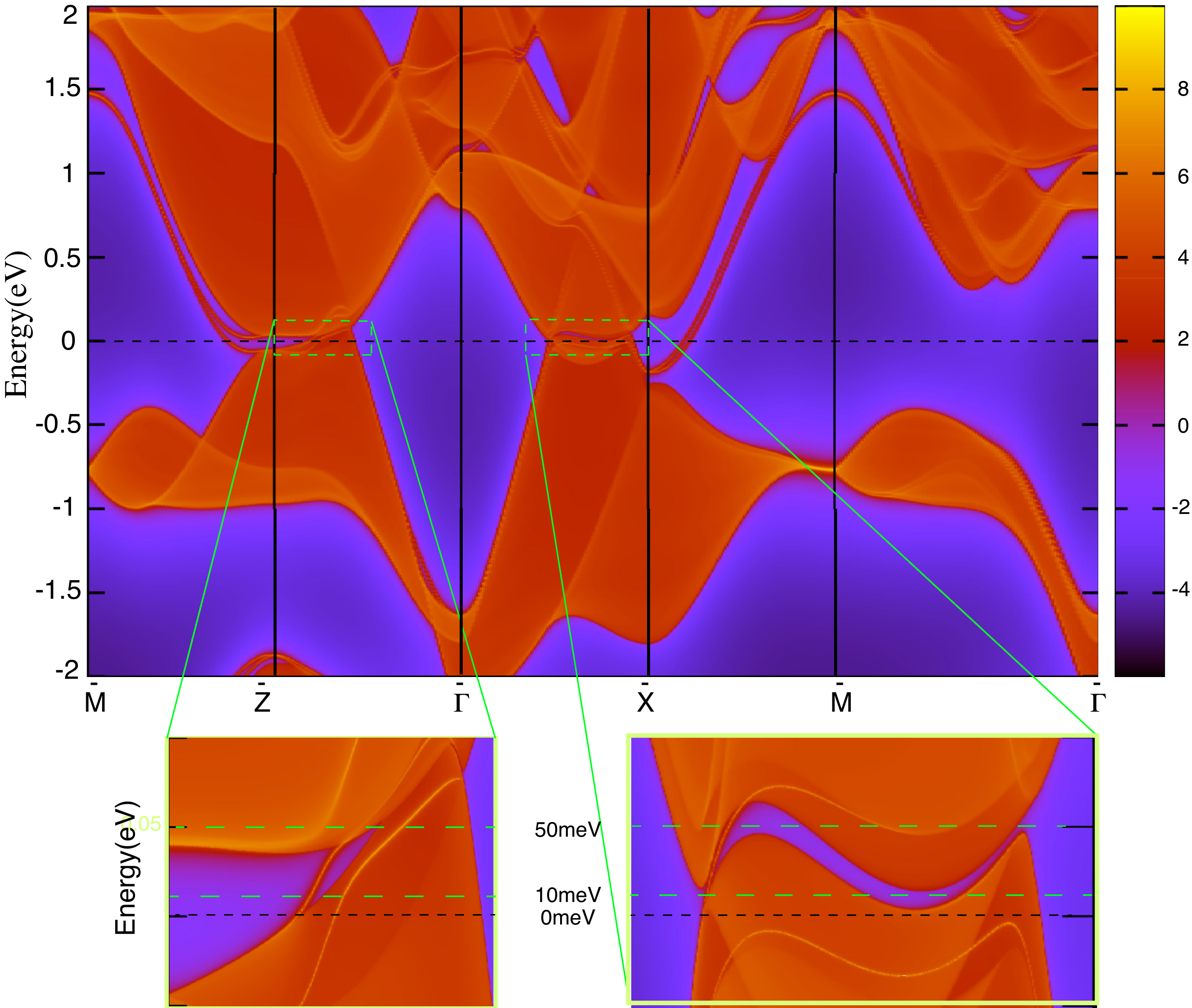}
\caption{
(Color online) ZrTe (100) surface state with its band structure weighted by the projection onto the one uint cell on surface layer. Part of the bands have been enlarged in the insets.
}
\label{ZrTe_100_Surface}
\end{figure}

\begin{figure}
\includegraphics[width=0.8\textwidth]{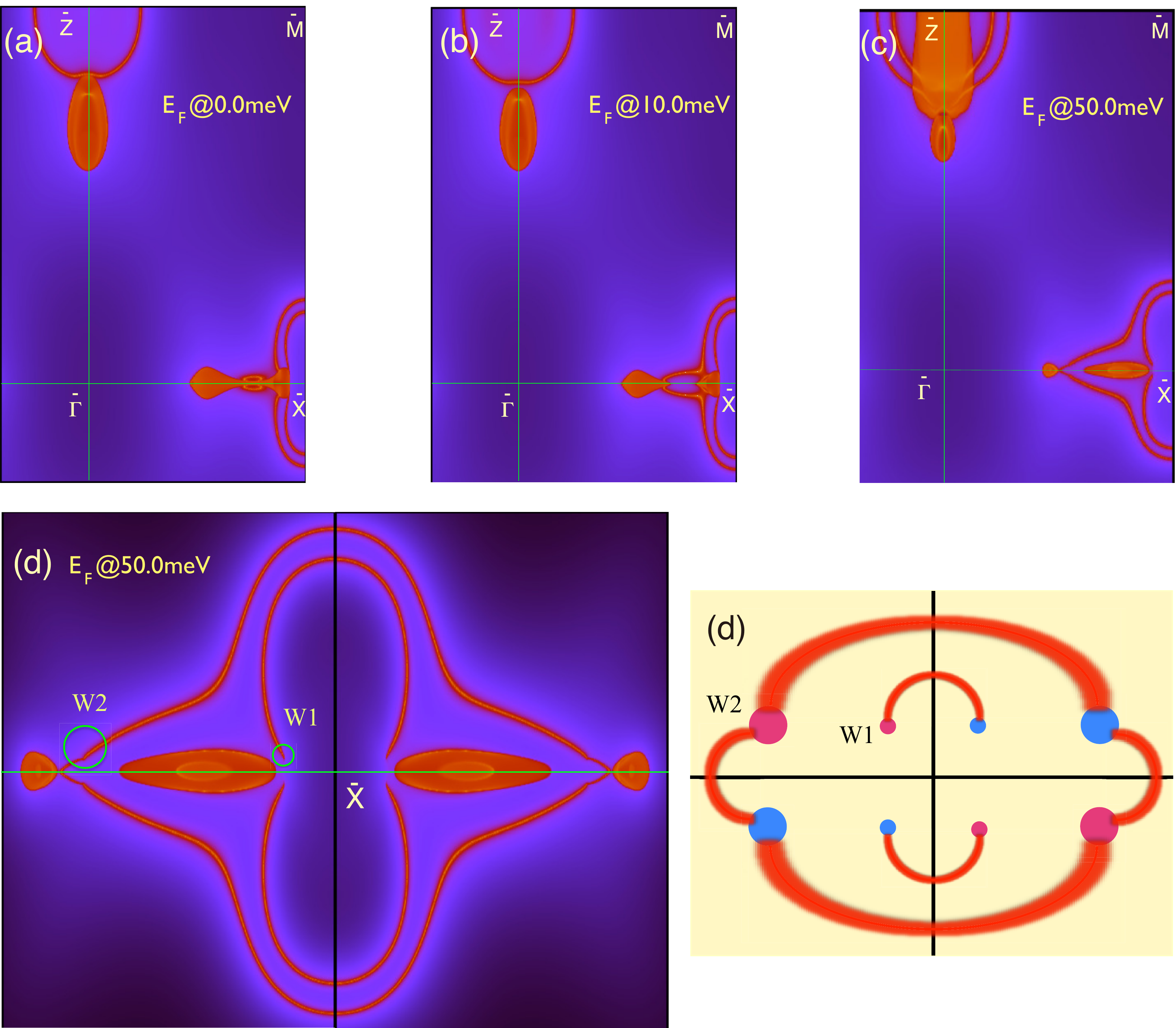}
\caption{
(Color online) Fermi surface of ZrTe (100) surface states. (a) chemical potential is set at 0 meV, (b) 10 meV and (c) 50 meV. Those around $\bar{X}$ in (c) is enlarged in (d). In (e)
the fermi arc pattern is schematically shown in (e).
}
\label{ZrTe_100_arc}
\end{figure}

{\it Surface states of ZrTe.}
The (100) surface states are shown in Fig.~\ref{ZrTe_100_Surface} with the surface band structure and Fig.~\ref{ZrTe_100_arc} with the Fermi surface 
at different chemical potentials, namely 0.0, 10.0 and 50.0 meV. The TDNPs are projected onto the $\bar{\Gamma}$-$\bar{Z}$ path, which is 
well separated from other surface states. The surface Dirac cone at $\bar{Z}$ is due to the $Z_2$=1 of the $k_z$=$\pi$ plane. Such Dirac cone is clearly
shown in the Fermi surface plot with chemical potential being 10.0 meV. When the Fermi level is set to at the Weyl nodes, i.e. 50 meV, 
the Fermi arcs connecting the projections of Weyl nodes becomes very clear since all the Weyl nodes and the arcs are well separated from the bulk states. 
As shown in Fig.~\ref{ZrTe_100_arc}(c) and (d), for each W2, the projection point of two identical Weyl nodes, there are two Fermi arcs connecting 
it with the other one of opposite chirality. For each W1, the projection point of one Weyl node, there are only one Fermi arc connecting it. 
The Fermi arc passes through the path $\bar{\Gamma}$-$\bar{X}$, where the $k_z$=0 mirror plane intersecting with the (100) surface. This passing
through is protected by the $Z_2$=1 of $k_z$=0 plane. This is similar as the case in TaAs,~\cite{TaAs_Weng,TaAs_arc} where the mirror Chern 
number plays the role. The schematic pattern of Fermi arc is drawn in Fig.~\ref{ZrTe_100_arc} (e).

\begin{figure}
\includegraphics[width=0.8\textwidth]{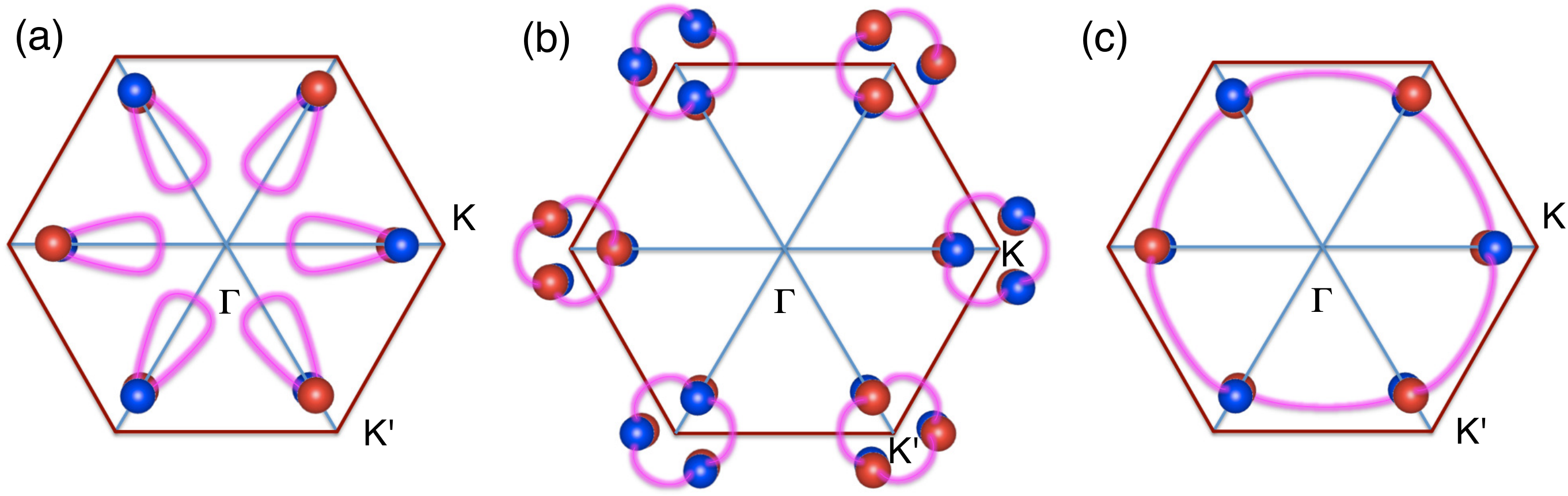}
\caption{(Color online) The possible connection pattern of Fermi arc on ZrTe (001) surface. (a) The arc connecting to superposed Weyl points. (b) The arcs form trimer. (c) The arcs form hexamer.
}
\label{ZrTe_arc_plot}
\end{figure}

Now we consider the (001)-surface states. On this surface, the two WP symmetric about $k_z=0$ are projected to the same point in the surface BZ. As they have opposite charges, each projection point contains one positive and one negative monopole charge, and there are six of such projection points. As we consider the configuration of the Fermi arcs, we note that their connection must follow the two constraints: (i) each projection point has and only has two arcs connected to it (ii) the connection pattern should preserve both $D_{3h}$ point group symmetry and time-reversal symmetry, which are the symmetries of the (001)-surface. Based on this analysis, we find three distinct configurations for the Fermi arcs. Fig.~\ref{ZrTe_arc_plot}(a) shows the case where each projection point connects to itself via one arc. Mark that this is equivalent to no arc at all since each lobe may shrink to a point. In Fig.~\ref{ZrTe_arc_plot}(b), the three projection points near and symmetric about $\bar{K}$ and $\bar{K'}$ connect to each other in a loop, forming a trimer. In Fig.~\ref{ZrTe_arc_plot}(c), all six projection points connect to each other in a large loop across the surface BZ, forming a hexamer. Which of the three cases may appear in a real sample depends on the details of the surface termination and is hence not related to the bulk topology.

\begin{figure}
\includegraphics[width=0.8\textwidth]{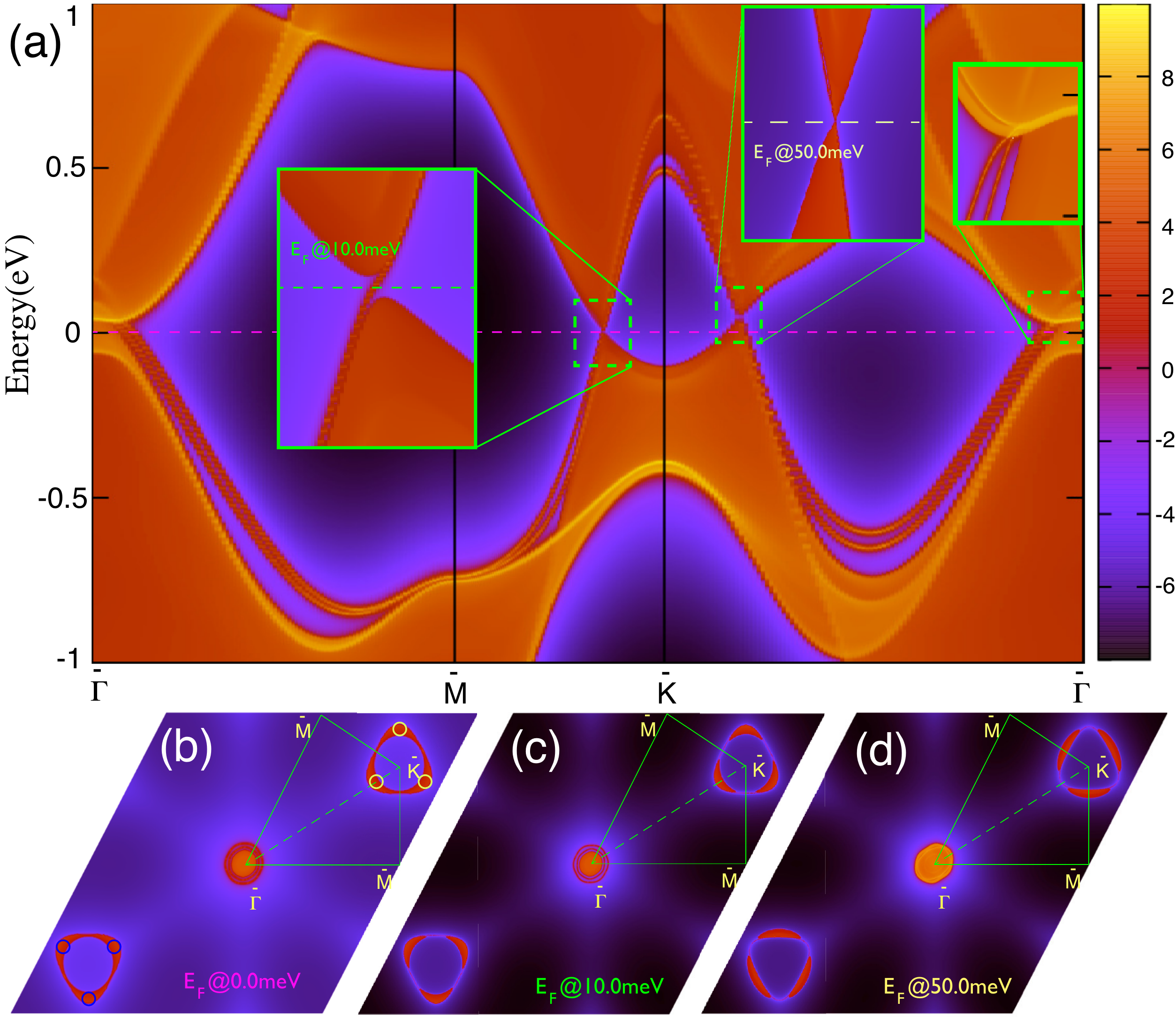}
\caption{(Color online) ZrTe (001) surface state. (a) The band structure weighted by the projection onto the one uint cell on top layer. (b), (c) and (d) are the Fermi surface at different chemical potential 0, 10 and 50 meV, respectively.
}
\label{ZrTe_001_Surface}
\end{figure}

The (001) surface states are shown in Fig.~\ref{ZrTe_001_Surface}. The bulk states around TDNPs are projected onto area around $\bar{\Gamma}$. The projections of Weyl nodes are indicated as circles in Fig.~\ref{ZrTe_001_Surface}(b). The band gap at 10 meV along $\bar{M}$-$\bar{K}$ is due to SOC and the in-gap surface bands are clearly shown in the inset of Fig.~\ref{ZrTe_001_Surface}(a), which contributes to the single thread like Fermi surface in Fig.~\ref{ZrTe_001_Surface}(c) crossing $\bar{M}$-$\bar{K}$. When the chemical potential is set at 50 meV, exactly passing through the Weyl nodes, the Fermi arc connecting to these their projections are clearly shown in Fig.~\ref{ZrTe_001_Surface}(d), where the $\bar{\Gamma}$-$\bar{K}$ path and the single thread like Fermi arcs intersect each other.

\begin{figure}
\includegraphics[width=0.7\textwidth]{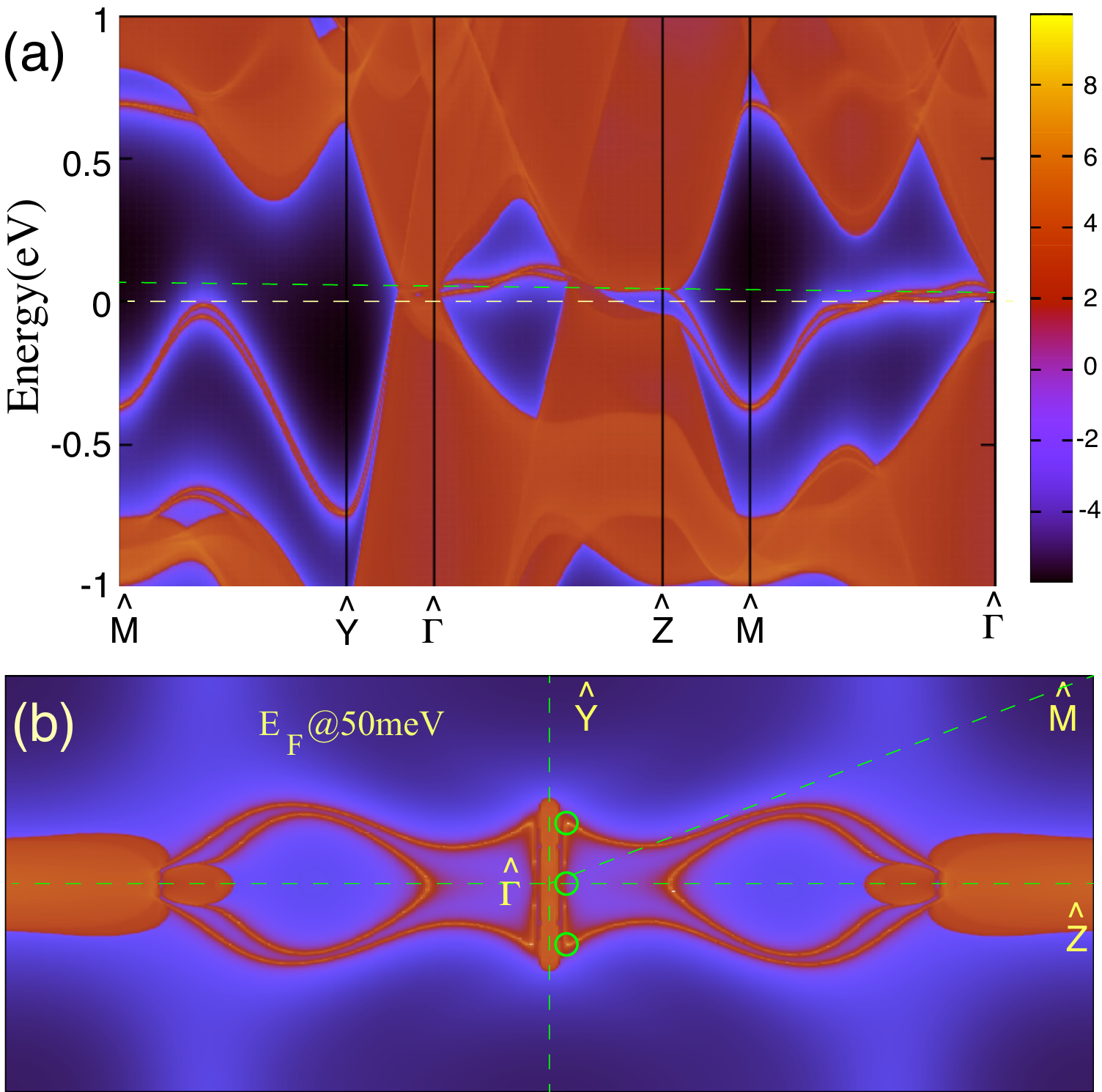}
\caption{(Color online) ZrTe (010) surface state. (a) The band structure weighted by the projection onto the one uint cell on top layer. (b) is the Fermi surface at 50 meV.
}
\label{ZrTe_010_Surface}
\end{figure}

The (010) surface states are shown in Fig.~\ref{ZrTe_010_Surface}. Similar to the (100) surface, the topologically nontrivial $k_z$=0 and $k_z$=$\pi$ planes will intersect the surface at $\hat{\Gamma}$-$\hat{Y}$ and $\hat{Z}$-$\hat{M}$, respectively. There are two surface Dirac cones centering $\hat{Y}$ and $\hat{M}$, respectively, seen in the surface band structure in Fig.~\ref{ZrTe_010_Surface}(a). Along $\hat{Y}$-$\hat{\Gamma}$ ($\hat{M}$-$\hat{Z}$), the two branches of the surface Dirac cone connects to the valence and conduction bulk states, respectively. The two Dirac cones are connected to each other along $\hat{M}$-$\hat{Y}$. When setting the chemical potential at 50 meV, the Fermi arcs connecting projections of Weyl nodes are clear, which are marked by the circles.

\section{Conclusions}
In this paper, we report the simultaneous presence of triply degenerate nodal points and Weyl points in ZrTe of WC-type. Due to the band inversion at K (and K') and the spin splitting, there emerge six pairs of Weyl points near K and K' in the 3D Brillouin zone. All Weyl points are related to each other via some symmetry operation from the point group $D_{3h}$, and hence are at the same energy level and have the same velocities. The surface states on the three low index surfaces, namely the (100)-, (010)- and (001)-surfaces, are calculated and shown. Using the (001)-surface states as example, we show that under the same symmetry constraints, there can be topologically different patterns of how Fermi arcs connect the Weyl points. The actual pattern in a real sample  depends on the conditions of the surface and hence may be tuned via surface engineerings.

\emph{Note added} Upon the completion of our work, we notice a new paper\cite{ZrTe_TDNP} that reports the triply degenerate nodal points in the same material, without studying the Weyl points therein.

\section{Acknowledgments}
We acknowledge the supports from National Natural Science Foundation of China (Grant Nos. 11274359 and 11422428),
the National 973 program of China (Grant No. 2013CB921700) and the ``Strategic Priority Research Program (B)"
of the Chinese Academy of Sciences (Grant No. XDB07020100). Partial of the calculations were preformed on TianHe-1(A), the National Supercomputer Center in Tianjin, China.

\bibliography{ZrTe_ref}

\end{document}